# Describing Strong Correlation with Block-Correlated Coupled Cluster Theory

Qingchun Wang,[†] Mingzhou Duan,[†] Enhua Xu,[‡] Jingxiang Zou,[†] Shuhua Li*,[†]

[†]Institute of Theoretical and Computational Chemistry, Key Laboratory of Mesoscopic Chemistry of Ministry of Education, School of Chemistry and Chemical Engineering, Nanjing University, Nanjing 210023, People's Republic of China.

[‡]Graduate School of Science, Technology, and Innovation, Kobe University, Nada-ku, Kobe 657-8501, Japan

**ABSTRACT**: A block-correlated coupled cluster (BCCC) method based on the generalized valence bond (GVB) wave function (GVB-BCCC in short) is proposed and implemented at the *ab initio* level, which represents an attractive multireference electronic structure method for strongly correlated systems. The GVB-BCCC method is demonstrated to provide accurate descriptions for multiple bond breaking in small molecules, although the GVB reference function is qualitatively wrong for the studied processes. For a challenging prototype of strongly correlated systems, tridecane with all 12 single C–C bonds at various distances, our calculations have shown that the GVB-BCCC2b method can provide highly comparable results as the density matrix renormalization group method for potential energy surfaces along simultaneous dissociation of all C–C bonds.

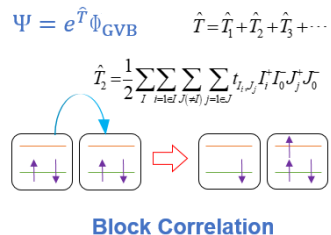

*Keywords: Generalized valence bond, Block correlation, Coupled cluster, Multireference, Strong correlation*

The unremitting pursuit of developing electronic structure theory is to accurately predict molecular geometries, energies and properties of various molecules and materials. Traditional single-reference (SR) electron correlation methods based on the Hartree–Fock (HF) reference have achieved great success for weakly correlated systems. Such SR methods include, for example, many-body perturbation theory[1] and coupled cluster (CC) method.[2] However, the accuracy of these methods decreases dramatically for strongly correlated systems (e.g, multiple-bond breaking, transition metal compounds), in which independent particle approximation is no longer suitable. Multireference (MR) electron correlation methods[3-7] have been developed to deal with strongly correlated systems. Those methods, which are based on the linear combination of many determinants as the reference function, have been demonstrated to provide satisfactory descriptions for small systems with strong correlation. However, the commonly-used reference function, the complete active space self-consistent-field (CASSCF) wave function[8] is only available for systems with small active spaces (i.e., 18 electrons in 18 active orbitals), due to the exponential increase of their computational cost with the size of the active space. Although the density matrix renormalization group (DMRG)[9-11] is established to be a powerful electronic structure method for strongly correlated linear molecules, it cannot provide satisfactory results for two- and three-dimensional systems. In addition, it is hard to incorporate dynamic correlation effectively within the DMRG framework to get quantitatively accurate results. Thus, the development of new MR electron structure methods, which are spin-pure, size-consistent, cost-effective, and accurate for a wide range of systems including both weakly and strongly correlated, remains a great challenge. In this Letter we describe an appealing method towards this goal.

For systems with strong correlation, the antisymmetric product of strongly orthogonal geminals (APSG)[12] and its special case, the perfect-pairing generalized valence bond (GVB-PP, denoted as GVB in short)[13] wave function is another well-known multiconfigurational reference function. Unlike the HF reference, the APSG and GVB reference functions can recover a small portion of static correlation and is capable of describing the correct dissociation of isolated single chemical bonds. In comparison with the CASSCF and DMRG reference functions, the APSG, especially GVB, reference functions are inexpensive and are computationally accessible for large systems. However, the APSG and GVB reference functions are not accurate enough for most systems, due to the lack of electron correlation between geminals, and between geminals and virtual orbitals.

The goal of the present work is to propose a block-correlated coupled cluster (BCCC) method based on the GVB reference (GVB-BCCC in short) for electronic structure calculations of strongly correlated molecules. In this method, a geminal is defined as a block, and the cluster operator is defined in terms of block electron states so that inter-geminal electron correlation is effectively treated. We have implemented GVB-BCCC methods at the *ab initio* level. This black-box GVB-BCCC method is spin-pure, size-consistent, cost-effective, and can provide accurate descriptions on the multi-bond dissociation processes, although the GVB reference is qualitatively wrong for such processes. For a relatively large tridecane molecule, we demonstrated that a simple GVB-BCCC method (GVB-BCCC2b) could provide accurate descriptions for simultaneous dissociation of its all 12 C–C bonds, which is much beyond the capability of the CASSCF method.

In second quantitation, the GVB wave function may be written as

$$\Phi_{\text{GVB}} = \prod_i \hat{a}^+_{i\uparrow} \hat{a}^+_{i\downarrow} \prod_\Lambda \hat{A}^+_\Lambda |\text{vac}\rangle, \quad \hat{A}^+_\Lambda = c^\Lambda_\chi \hat{a}^+_{\chi\uparrow} \hat{a}^+_{\chi\downarrow} + c^\Lambda_\gamma \hat{a}^+_{\gamma\uparrow} \hat{a}^+_{\gamma\downarrow}, \quad (1)$$

where $\hat{a}^+_{i\uparrow}$ ($\hat{a}^+_{i\downarrow}$) represents the creation operator of a spin-up (spin-down) electron in core orbitals, which can be regarded as inactive orbitals, and $\hat{A}_\Lambda^\dagger$ stands for the creation operator of a singlet pair (or geminal) of electrons in two orbitals, $\chi$ and $\gamma$, which corresponds to the first and second natural orbital (NO) in the $\Lambda$th geminal. Here, a geminal is a kind of strongly orthogonal singlet two-electron function generalized from one-electron functions (orbitals), and is the linear combination of two determinants. In the GVB wave function, two orbitals in each geminal subspace correspond to a bonding orbital, and its antibonding orbital localized over a chemical bond. A geminal can provide a good description for strong correlation of an electron pair in a chemical bond. For an open-shell molecule, there are also singly occupied orbitals in the GVB wave function (neglected in Eq. (1)). By construction, the GVB wave function is of a multiconfigurational character and includes some highly excited determinants (relative to the corresponding HF determinant). Thus, the GVB wave function is capable of describing the exact ground state of the supramolecular composed of any number of noninteracting electron pairs and is also size-extensive, and size-consistent. Recent development of the GVB method has allowed black-box GVB calculations to become routine for systems with hundreds of electrons.[14]

The main deficiency of the GVB wave function is the lack of electron correlation between geminals, and between geminals and virtual orbitals that prevents the GVB method from being a quantitatively reliable electronic structure method. To include the remaining electron correlation mentioned above, different GVB-based approaches[15-18] or APSG-based approaches[19-20] have been proposed. However, none of these approaches can provide satisfactory accuracy for strongly correlated systems.

Keeping in mind that the ansatz of the traditional CC cluster method is powerful in describing dynamic electron correlation, one may wonder whether it is possible to develop an alternative CC framework for strongly correlated systems. In fact, such a CC framework, called block-correlated CC (BCCC) method, was proposed by our group in 2004.[21] In BCCC, all spin orbitals are divided into many blocks, and a block is defined to consist of two or more localized spin orbitals, characterized by capital letters $I$, $J$, $A$, $B$, etc. A block state is defined as a many-electron state in the Fock space. For a given block $I$, its $i$th electronic state is denoted by $|I_i\rangle$, which can be created by a block-state creation operator $I_i^+$ acting on the vacuum state,

$$|I_i\rangle = I_i^+|\text{vac}\rangle. \quad (2)$$

Here $I_i^+$ is a linear combination of the product of the creation operators of all spin orbitals in block $I$. $\{|I_i\rangle, i=0,1,\cdots,n_I-1\}$ ($n_I$ is the total number of many-electron states) represents the complete set of many-electron states for this block, and all electronic states in a block must be orthogonal to each other. For convenience, $|I_0\rangle$ is defined to be the ground state of this block, and all of the other electronic states are called as excited block states. The reference function of the whole system in the BCCC method is defined as the tensor product of the ground states of all blocks. For example, the reference function may be written as

$$\Phi_0 = |1_0\rangle|2_0\rangle\cdots|N_0\rangle. \quad (3)$$

Relative to the reference function, all other configuration functions can be defined as block-correlated configuration functions. For example, a doubly correlated configuration function can be written as

$$\Phi_{A_aB_b} = A_a^+A_0^-B_b^+B_0^-\Phi_0. \quad (4)$$

Here the ground states of two blocks, A and B, are replaced by the corresponding excited block states ($|A_a\rangle$, $|B_b\rangle$) and other blocks remain in their ground states.

Within the BCCC framework, the exact ground-state wave function can be formulated as

$$\Psi = e^{\hat{T}}\Phi_0, \quad (5)$$

where

$$e^{\hat{T}} = 1+\hat{T}+\frac{\hat{T}^2}{2!}+\frac{\hat{T}^3}{3!}+\cdots = \sum_{k=0}^{\infty}\frac{\hat{T}^k}{k!}. \quad (6)$$

Here the cluster operator $\hat{T}$ is defined as the sum of connected $n$-block correlation operators,

$$\hat{T} = \hat{T}_1+\hat{T}_2+\hat{T}_3+\cdots+\hat{T}_n, \quad (7)$$

$$\hat{T}_1 = \sum_I\sum_{i=1}^{}t_{I_i}I_i^+I_0^-, \quad (8)$$

$$\hat{T}_2 = \frac{1}{2}\sum_I\sum_{i=1\in I}\sum_{J(\neq I)}\sum_{j=1\in J}t_{I_i,J_j}I_i^+I_0^-J_j^+J_0^- \quad (9)$$

$$\hat{T}_3 = \frac{1}{6}\sum_I\sum_{i=1}\sum_{J(\neq I)}\sum_{j=1}\sum_{K(\neq I,J)}\sum_{k=1}t_{I_iJ_jK_k}I_i^+I_0^-J_j^+J_0^-K_k^+K_0^-, \quad (10)$$

where coefficients $t_{I_i}$, $t_{I_i,J_j}$ and $t_{I_iJ_jK_k}$ are called single, double, triple correlation amplitudes, respectively. It should be pointed out that the 2-block correlation operator acting on the reference function will generate a linear combination of all doubly correlated configuration functions, as shown in Figure 1. The BCCC ansatz can be considered as an extension of the traditional CC ansatz, in which the cluster operators and the reference function here are defined in terms of block states instead of orbitals in the traditional CC method. It was expected that this BCCC method represents a very attractive solution to the electronic structure of strongly correlated systems. However, the implementation of the BCCC method with a reasonable reference function at the *ab initio* level is a very difficult task, and thus has never been reported.

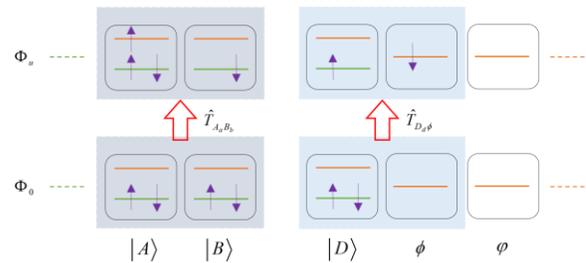

**Figure 1**. An illustrative picture of doubly correlated configuration functions generated by the 2-block correlation operator between two geminals ($\hat{T}_{A_aB_b}$) and between a geminal and a virtual orbital ($\hat{T}_{A_a\phi}$).

Now we can outline the formulation of the GVB-BCCC method proposed in this Letter. The GVB-BCCC ground-state wave function is expressed as

$$\Psi = e^{\hat{T}}\Phi_{\text{GVB}}. \quad (11)$$

In GVB-BCCC, a block may be defined as a geminal (containing two spatial orbitals) or a spatial orbital. For example, if a geminal is taken as a block, there are 16 many-electron states (a ground state and 15 excited states) in the Fock space, as shown in Table 1. Truncating the cluster operator to a certain $n$-block correlation level, an approximate GVB-BCCC method, denoted as GVB-BCCC$n$, is defined. In this work, GVB-BCCC2 ($\hat{T} \approx \hat{T}_1 + \hat{T}_2$) and GVB-BCCC3 ($\hat{T} \approx \hat{T}_1 + \hat{T}_2 + \hat{T}_3$) have been implemented for small and medium-sized molecules, in which inter-geminal electron correlation has been taken into account. It is worth mentioning that correlation between geminals and virtual orbitals is not considered here, but could be added without much difficulty in the near future, as was done in our previous CAS-BCCC method.[22] For relatively large molecules, an approximate GVB-BCCC2, called GVB-BCCC2b ($\hat{T} \approx \hat{T}_2$), has been implemented.

**Table 1. Expressions for 16 Electronic States within a Geminal $\Lambda$ (the Number of Electron ($n_e$) and the Z-component of the Total Spin ($S_z$) are Given for Each State).**

| $\Lambda_\lambda$ | Expression[a] | $n_e$ | $S_z$ |
|---|---|---|---|
| $\lambda=0$ | $(c_\chi \chi_\alpha^+ \chi_\beta^+ + c_\gamma \gamma_\alpha^+ \gamma_\beta^+)\|\text{vac}\rangle$ [b] | 2 | 0 |
| $\lambda=1$ | $(-c_\gamma \chi_\alpha^+ \chi_\beta^+ + c_\chi \gamma_\alpha^+ \gamma_\beta^+)\|\text{vac}\rangle$ | 2 | 0 |
| $\lambda=2$ | $(\frac{1}{\sqrt{2}} \chi_\alpha^+ \gamma_\beta^+ - \frac{1}{\sqrt{2}} \chi_\beta^+ \gamma_\alpha^+)\|\text{vac}\rangle$ | 2 | 0 |
| $\lambda=3$ | $(\frac{1}{\sqrt{2}} \chi_\alpha^+ \gamma_\beta^+ + \frac{1}{\sqrt{2}} \chi_\beta^+ \gamma_\alpha^+)\|\text{vac}\rangle$ | 2 | 0 |
| $\lambda=4$ | $\chi_\alpha^+ \gamma_\alpha^+ \|\text{vac}\rangle$ | 2 | +1 |
| $\lambda=5$ | $\chi_\beta^+ \gamma_\beta^+ \|\text{vac}\rangle$ | 2 | −1 |
| $\lambda=6$ | $\|\text{vac}\rangle$ | 0 | 0 |
| $\lambda=7$ | $\chi_\alpha^+ \|\text{vac}\rangle$ | 1 | +1/2 |
| $\lambda=8$ | $\gamma_\alpha^+ \|\text{vac}\rangle$ | 1 | +1/2 |
| $\lambda=9$ | $\chi_\beta^+ \|\text{vac}\rangle$ | 1 | −1/2 |
| $\lambda=10$ | $\gamma_\beta^+ \|\text{vac}\rangle$ | 1 | −1/2 |
| $\lambda=11$ | $\chi_\alpha^+ \chi_\beta^+ \gamma_\alpha^+ \|\text{vac}\rangle$ | 3 | +1/2 |
| $\lambda=12$ | $\chi_\alpha^+ \gamma_\alpha^+ \gamma_\beta^+ \|\text{vac}\rangle$ | 3 | +1/2 |
| $\lambda=13$ | $\chi_\alpha^+ \chi_\beta^+ \gamma_\beta^+ \|\text{vac}\rangle$ | 3 | −1/2 |
| $\lambda=14$ | $\chi_\beta^+ \gamma_\alpha^+ \gamma_\beta^+ \|\text{vac}\rangle$ | 3 | −1/2 |
| $\lambda=15$ | $\chi_\alpha^+ \chi_\beta^+ \gamma_\alpha^+ \gamma_\beta^+ \|\text{vac}\rangle$ | 4 | 0 |

[a] $\chi$ and $\gamma$ stand for the first and second natural orbitals in the $\Lambda$th geminal respectively. [b] $c_\chi$ and $c_\gamma$ are CI coefficients of two configurations from GVB calculations.

The procedure of determining the amplitudes and the ground-state energy in GVB-BCCC theory is similar to that in the traditional CC method. Take the GVB-BCCC2b method as an example. In this case, the cluster operator $\hat{T}$ is truncated as $\hat{T} \approx \hat{T}_2$. By projecting the electronic Schrödinger equation to the GVB reference function, one can obtain the energy equation,

$$\langle \Phi_{\text{GVB}} | \hat{H} | \Psi_{\text{GVB-BCCC2b}} \rangle = \langle \Phi_{\text{GVB}} | E | \Psi_{\text{GVB-BCCC2b}} \rangle = E . \quad (12)$$

The ground-state GVB-BCCC2b energy will depend on doubly correlated amplitudes ($t_{A_a B_b}$). Similarly, one may obtain the expressions for determining all amplitudes by left-projecting the Schrödinger equation on doubly correlated configuration functions respectively,

$$\langle \Phi_{A_a B_b} | \hat{H} | \Psi_{\text{GVB-BCCC2b}} \rangle = E \langle \Phi_{A_a B_b} | \Psi_{\text{GVB-BCCC2b}} \rangle = E \times t_{A_a B_b} . \quad (13)$$

By substituting the GVB-BCCC2b wave function into Eq. (13), we can obtain:

$$\langle \Phi_{A_a B_b} | \hat{H} | \Psi_{\text{GVB-BCCC2b}} \rangle = \langle \Phi_{A_a B_b} | \hat{H} [1 + \hat{T}_2 + \frac{1}{2} \hat{T}_2^2 + \frac{1}{6} \hat{T}_2^3] \Phi_{\text{GVB}} \rangle \\ = E \times t_{A_a B_b} \quad (14)$$

It should be mentioned that the expression of Eq. (14) is very complicated, and will be omitted for simplicity. By substituting the expressions of the ground-state energy, one can obtain a set of coupled nonlinear equations, which can be solved by using iterative technique. Once the amplitudes ($t_{A_a B_b}$) are known, the GVB-BCCC2b ground-state energy can be calculated.

It is beneficial for readers to know how to derive the expressions for the terms in the left-hand side of Eq. (14). Remember that the Hamiltonian operator can be rewritten as the sum of all single-, two-, three-, and four-block terms,

$$\hat{H} = \sum_P \hat{H}_P + \frac{1}{2} \sum_{PQ} \hat{H}_{PQ} + \frac{1}{3!} \sum_{PQR} \hat{H}_{PQR} + \frac{1}{4!} \sum_{PQRS} \hat{H}_{PQRS} . \quad (15)$$

Here capital letters $P$, $Q$, $R$, and $S$ stand for different block indices. For example, one of the two-block terms, $\hat{H}_{PQ}$, can be explicitly formulated as

$$\hat{H}_{PQ} = \sum_{\substack{p \in P, \\ q \in Q}} \sum_a \langle p | \hat{h} | q \rangle p_a^+ q_a^- + \sum_{\substack{p \in Q, \\ q \in P}} \sum_a \langle p | \hat{h} | q \rangle p_a^+ q_a^- \\ + \frac{1}{4} \sum_{\substack{p \in P, \\ qrs \in Q}} \sum_{a,b} \langle pq \| rs \rangle p_a^+ q_b^+ s_b^- r_a^- \\ + \frac{1}{4} \sum_{\substack{p \in Q, \\ qrs \in P}} \sum_{a,b} \langle pq \| rs \rangle p_a^+ q_b^+ s_b^- r_a^- + \cdots \quad (16)$$

$$\langle p | \hat{h} | q \rangle = \int d\mathbf{r}_1 \phi_p^*(\mathbf{r}_1) \hat{h}(\mathbf{r}_1) \phi_q(\mathbf{r}_1) . \quad (17)$$

$$\langle pq \| rs \rangle = \langle pq | \hat{g} | rs \rangle - \langle pq | \hat{g} | sr \rangle \\ = \int d\mathbf{r}_1 d\mathbf{r}_2 \phi_p^*(\mathbf{r}_1) \phi_q^*(\mathbf{r}_2) r_{12}^{-1} (1 - \hat{P}_{12}) \phi_r(\mathbf{r}_1) \phi_s(\mathbf{r}_2) \quad (18)$$

where $p$, $q$, $r$ and $s$ represent spatial molecular orbital (MO) indices, $a$ and $b$ represent the two spin states of $\alpha$ (up) or $\beta$ (down), $\hat{P}_{12}$ is an operator which interchanges the coordinates of electron 1 and 2, $\langle p | \hat{h} | q \rangle$ is a standard MO-based one-electron integral, while $\langle pq | \hat{g} | rs \rangle$ is a standard MO-based two-electron integral. Therefore, the Hamiltonian matrix elements between any two configuration functions $\Phi_u$ and $\Phi_v$ can be calculated as follows:

$$\langle \Phi_u | \hat{H} | \Phi_v \rangle = \langle \Phi_u | \sum_P \hat{H}_P + \frac{1}{2} \sum_{PQ} \hat{H}_{PQ} + \frac{1}{3!} \sum_{PQR} \hat{H}_{PQR} \\ + \frac{1}{4!} \sum_{PQRS} \hat{H}_{PQRS} | \Phi_v \rangle \quad (19)$$

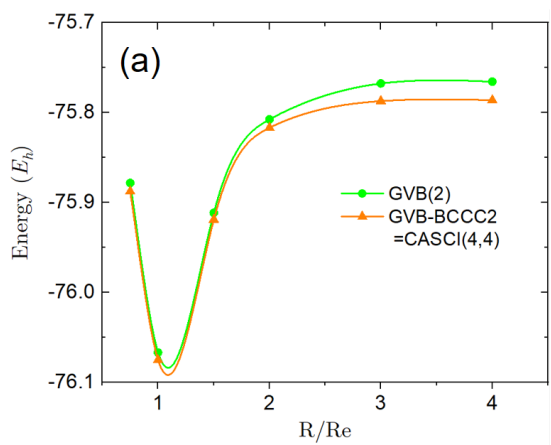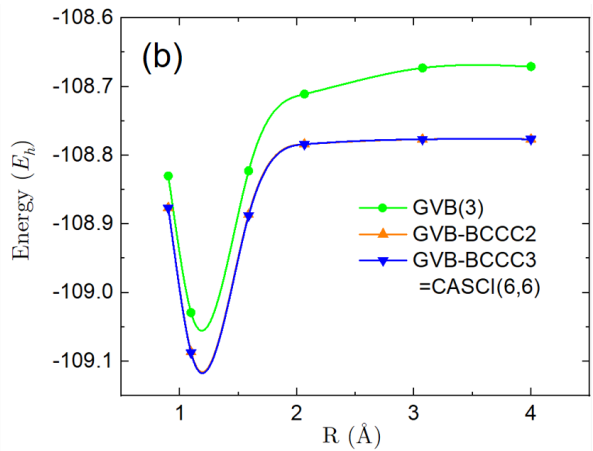

**Figure 2**. The ground-state potential energy surfaces for (a) symmetric dissociation in $H_2O$ (bond angle 110.6° and Re = 0.97551 Å) at the Cartesian cc-pVDZ basis set; and (b) triple bond dissociation in $N_2$ at the Cartesian cc-pVDZ basis set.

$$\langle \Phi_u | \hat{H}_{PQ} | \Phi_v \rangle$$
$$= \sum_{\substack{p \in P, \\ q \in Q}} \sum_a \langle p|h|q \rangle \hat{\mathcal{P}}(q_a^-, P_i) \langle P_{i'} | p_a^+ | P_i \rangle \langle Q_{j'} | q_a^- | Q_j \rangle + \ldots$$
$$+ \frac{1}{4} \sum_{\substack{p \in P, \\ qrs \in Q}} \sum_{a,b} \langle pq||rs \rangle \hat{\mathcal{P}}(q_b^+ s_b^- r_a^-, P_i) \langle P_{i'} | p_a^+ | P_i \rangle \langle Q_{j'} | q_b^+ s_b^- r_a^- | Q_j \rangle$$
$$+ \ldots \qquad (20)$$

where $\hat{\mathcal{P}}(\hat{O}, P_i)$ is +1 or –1 associated with the operation $\hat{O}|P_i\rangle \to |P_i\rangle \hat{O}$, depending on the number of particles in the block-state $P_i$. Once the matrix representations of some operators (such as $p_a^+$ and $q_b^+ s_b^- r_a^-$) within all blocks are calculated, the Hamiltonian matrix elements between any two configuration functions can be calculated as shown above. By introducing a number of the intermediate arrays, we can demonstrate that the computational cost of the GVB-BCCC2b (or GVB-BCCC2) method scales with the system size as $O(N^4)$, and that of the GVB-BCCC3 method scales as $O(N^5)$. Thus, with optimized programs, GVB-BCCC calculations may be computationally manageable for quite large systems.

With the procedure described above, we have implemented GVB-BCCC2 and GVB-BCCC3 methods for molecules with a few pairs of electrons and GVB-BCCC2b for relatively large systems. For systems under study, HF and GVB calculations in this Letter are performed with the Gaussian 16 package[23] and a modified GAMESS,[24] respectively. CASCI or DMRG calculations are carried out with the PySCF program.[25] It is well known that the GVB reference function is size-consistent. Naturally, the GVB-BCCC method truncated at any level should be size-consistent. We have taken two linear $H_6$ molecules (with the distance between two neighboring H atoms being 2.0 Å) separated by 50 Å as an example to demonstrate this point. The GVB-BCCC2 energy of the system is –5.745348 $E_h$ at the STO-6G basis set, which is exactly the sum of the energies (–2.872674 $E_h$) of two linear $H_6$ molecules. As expected, the GVB-BCCC2 or GVB-BCCC3 method exactly preserves size-consistency.

A nice feature of the GVB-BCCC method is that it can provide exact descriptions on the double bond dissociation in $H_2O$ at the GVB-BCCC2 level, and on the triple bond dissociation in $N_2$ at the GVB-BCCC3 level. One can see from Figure 2 that the GVB-BCCC2 result is equivalent to the CASCI (4,4) value (based on GVB orbitals) for the symmetric dissociation in $H_2O$, and the GVB-BCCC3 result is the same as the CASCI(6,6) result (based on GVB orbitals) for triple bond dissociation in $N_2$, although the GVB reference function is very poor when the bond approaches the dissociation limit.

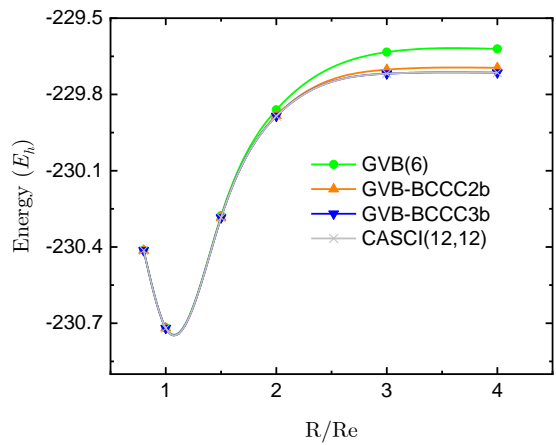

**Figure 3**. Potential energy curves for simultaneous breaking of all C–H bonds in benzene (Re = 1.08557 Å) with the 6-31G basis set.

The symmetric dissociation of all 6 C-H bonds in benzene is a typical strong correlated system, and we have tested the performance of the GVB-BCCC methods for this system. At the 6-31G basis set, the CASCI method based on the active space (12,12) is required to give accurate descriptions for various geometries with difference C–H bond distances (other structural parameters remain unchanged at the B3LYP/6-31G-optimzied values). The potential energy surfaces obtained from GVB-BCCC2b and GVB-BCCC3b calculations are plotted in Figure 3, together with the CASCI(12,12) results. We can see that the GVB method has a poor performance at large C-H distances, and the maximum absolute deviation with respect to the CASCI(12,12) result is about 66.7 kcal/mol. However, GVB-BCCC2b captures 70% of the correlation energy (the difference between the CASCI and GVB energies may be defined as the correlation energy), and the GVB-BCCC3b method provides about 90% of the total correlation energy. Along the whole potential energy surface, the GVB-BCCC3b curve is quite close to the CASCI(12,12) profile, with the maximum absolute deviation of about 7.8 kcal/mol.

Polyacenes, which exhibit strong correlation in the π space, were also chosen to verify the applicability of the GVB-BCCC method. For several polyacenes, we have calculated their π-electron correlation energies at the 6-31G basis set. Here we perform GVB, GVB-BCCC, and DMRG calculations for all π-electrons only. The DMRG results are taken as the reference data since the CASCI method is not available for these molecules. As shown in Table 2, GVB-BCCC2b and GVB-BCCC2 can provide 60% of the total correlation energy, and the small difference between them shows that GVB-BCCC2b is a very good approximation to the GVB-BCCC2 method. When the triple block correlation is included, one can see that more than 90% of the total correlation energy can be obtained at the GVB-BCCC3b and GVB-BCCC3 level. The maximum absolute deviation is only 4.5 kcal/mol. These results demonstrate that the three-geminal correlation plays an important role for polyacenes.

**Table 2. Deviations of the Ground-State Energies (in kcal/mol) Calculated with GVB, GVB-BCCC2b and GVB-BCCC3b Methods at the 6-31G Basis Set for Several Polyacenes. The DMRG Ground-State Energies (Based on GVB Orbitals and M=1000) are Provided as the Reference Data.** [a]

| Polyacenes | $N$[b] | GVB | GVB-BCCC2b | GVB-BCCC2 | GVB-BCCC3b | GVB-BCCC3 | DMRG ($E_h$) |
|---|---|---|---|---|---|---|---|
| Anthracene ($C_{14}H_{10}$) | 14 | 38.68 | 15.93 | 15.57 | 2.73 | 2.61 | –535.98648 |
| Phenanthrene ($C_{14}H_{10}$) | 14 | 38.90 | 14.37 | 14.08 | 1.69 | 1.60 | –535.99481 |
| Pyrene ($C_{16}H_{10}$) | 16 | 48.63 | 21.39 | 20.96 | 3.91 | 3.77 | –611.74912 |
| Tetracene ($C_{18}H_{12}$) | 18 | 50.13 | 21.29 | 20.75 | 4.41 | 4.33 | –688.62580 |
| Tetraphene ($C_{18}H_{12}$) | 18 | 51.35 | 20.43 | 19.99 | 2.91 | 2.84 | –688.63888 |
| Chrysene ($C_{18}H_{12}$) | 18 | 49.91 | 18.89 | 18.49 | 2.55 | 2.48 | –688.64221 |
| MAE | | 51.35 | 21.39 | 20.96 | 4.41 | 4.33 | |
| NPE | | 12.66 | 7.02 | 6.88 | 2.71 | 2.73 | |

[a]The equilibrium geometries of polyacenes were optimized at the B3LYP/6-31G level. [b]The number of π electrons.

A more challenging system is tridecane, in which all C–C bonds are stretched simultaneously to the dissociation limit, and all bond angles and C–H bond lengths remain unchanged at their optimized values at the B3LYP/cc-pVDZ level. To describe the system accurately, one should require an active space (24,24) (containing a pair of orbitals for each of 12 C–C bonds) for the correlation calculation (1s orbital of each C atom and orbitals of all C–H bonds may be frozen here). This challenging system is much beyond the capability of many existing MR electronic structure methods based on the CASSCF and CASCI methods. Fortunately, the DMRG(24,24) method is able to provide the nearly exact CASCI(24,24) ground-state energy, and thus the DMRG(24,24) energies will be taken as the reference values. For this system, we can only do GVB-BCCC2b calculations. For this system at various C–C bond distances, the deviations of GVB and GVB-BCCC2b ground-state energies from the reference values are collected in Table 3. As shown in Table 3, one can see that the GVB method has a very poor performance. The maximum absolute deviation is as large as 119.1 kcal/mol when the C–C bond length is 4.0 times the equilibrium value (Re). On the contrary, the GVB-BCCC2b method is demonstrated to provide quite accurate results, with the largest deviation being only 5.3 kcal/mol at the bond length of 2.0 Re. When the length of C–C bond is 4.0 Re, there is only a deviation of 0.8 kcal/mol. Thus, GVB-BCCC2b is capable of accurately describing simultaneous dissociation of 12 C–C bonds in tridecane. This result indicates that two-geminal correlation play a dominant role for the electron correlation in this system. To better understand this point, a pair of GVB orbitals is shown in Figure 4. One can see clearly that each pair of orbitals (one bonding, and one antibonding) is very localized on a single C–C bond for tridecane. This picture explains why GVB-BCCC2b works well for this system. One can expect that GVB-BCCC2b should work well for other systems with similar electronic structures.

**Table 3. Deviations of the Ground-State Energies (in kcal/mol) Calculated with GVB and GVB-BCCC2b Methods at the Cartesian cc-pVDZ Basis Set for Simultaneous Bond Dissociation of All C–C Single Bonds in a Tridecane Molecule. The DMRG(24,24) Ground-State Energies (Based on GVB Orbitals and M=1000) are Provided as the Reference Data.** [a]

| R/Re | GVB(12) | GVB-BCCC2b | DMRG(24,24) ($E_h$) |
|---|---|---|---|
| 0.8 | 15.65 | 1.79 | –507.626761 |
| 1 | 22.48 | 2.47 | –508.860371 |
| 1.5 | 43.03 | 4.22 | –507.800702 |
| 2 | 76.75 | 5.28 | –507.179142 |
| 3 | 118.40 | 1.03 | –507.098237 |
| 4 | 119.14 | 0.83 | –507.098002 |
| MAE | 119.14 | 5.28 | |
| NPE | 103.48 | 4.45 | |

[a]The equilibrium geometry of tridecane was optimized at the B3LYP/cc-pVDZ level.

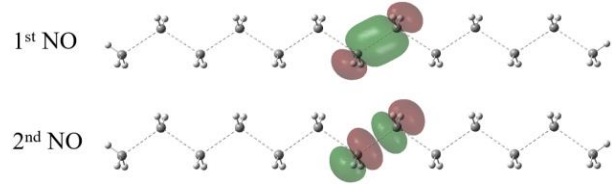

**Figure 4.** A pair of GVB orbitals localized at a given C–C bond in tridecane (R = 2.0 Re).

To conclude, we have proposed a GVB-based block correlated coupled cluster method and reported the first implementation of this method at the *ab initio* level. This black-box MR method is capable of providing accurate electronic structures for strongly correlated systems. We have demonstrated that GVB-BCCC2 can provide accurate descriptions on the double bond dissociation in $H_2O$ and GVB-BCCC3 can offer accurate descriptions on the triple bond dissociation in $N_2$. For polyacenes and bond dissociation in benzene, we have demonstrated that the GVB-BCCC3b method can provide quite accurate descriptions. For tridecane with all 12 single C–C bonds at various distances, our calculations have shown that the GVB-BCCC2b method can provide highly comparable results as the DMRG method from the weakly correlated region to the highly correlated region (bond dissociation limit). An advantage of the GVB-BCCC method over the DMRG method is that the dynamical electron correlation between geminals and virtual orbitals neglected in this work can be readily incorporated via adding the corresponding cluster operators, which is the topic of our future study. The GVB-BCCC method is expected to become a potentially powerful theoretical method for electronic structure calculations of strongly correlated systems.

*E-mail: shuhua@nju.edu.cn.

ACKNOWLEDGMENTS

This work was supported by the National Natural Science Foundation of China (Nos. 21833002 and 21673110). And all calculations were performed on the IBM Blade cluster system in the High Performance Computing Center of Nanjing University.

## REFERENCES

(1) Møller, C.; Plesset, M. S. Note on an Approximation Treatment for Many-Electron Systems. *Phys. Rev.* **1934,** *46*, 618-622.
(2) Bartlett, R. J.; Musiał, M. Coupled-cluster theory in quantum chemistry. *Rev. Mod. Phys.* **2007,** *79*, 291-352.
(3) Offermann, R. Degenerate many fermion theory in expS form. *Nucl. Phys. A* **1976,** *273*, 368-382.
(4) Soliverez, C. E. General theory of effective Hamiltonians. *Phys. Rev. A* **1981,** *24*, 4-9.
(5) Jeziorski, B.; Paldus, J. Spin‐adapted multireference coupled‐cluster approach: Linear approximation for two closed‐shell‐type reference configurations. *J. Chem. Phys.* **1988,** *88*, 5673-5687.
(6) Hubac, I. I.; Neogrady, P. Size-consistent Brillouin-Wigner perturbation theory with an exponentially parametrized wave function: Brillouin-Wigner coupled-cluster theory. *Phys. Rev. A* **1994,** *50*, 4558-4564.
(7) Mahapatra, U. S.; Datta, B.; Mukherjee, D. A size-consistent state-specific multireference coupled cluster theory: Formal developments and molecular applications. *J. Chem. Phys.* **1999,** *110*, 6171-6188.
(8) O., R. B. The Complete Active Space Self‐Consistent Field Method and its Applications in Electronic Structure Calculations. In *Advances in Chemical Physics*, 1987; pp 399-445.
(9) White, S. R. Density matrix formulation for quantum renormalization groups. *Phys. Rev. Lett.* **1992,** *69*, 2863-2866.
(10) Chan, G. K.; Sharma, S. The density matrix renormalization group in quantum chemistry. *Annu. Rev. Phys. Chem.* **2011,** *62*, 465-81.
(11) Szalay, S., et al. Tensor product methods and entanglement optimization for ab initio quantum chemistry. *Int. J. Quantum. Chem.* **2015,** *115*, 1342-1391.
(12) Parks, J. M.; Parr, R. G. Theory of Separated Electron Pairs. *J. Chem. Phys.* **1958,** *28*, 335-345.
(13) Bobrowicz, F. W.; Goddard, W. A. The Self-Consistent Field Equations for Generalized Valence Bond and Open-Shell Hartree—Fock Wave Functions. In *Methods of Electronic Structure Theory*, Schaefer, H. F., Ed. Springer US: Boston, MA, 1977; pp 79-127.
(14) Wang, Q., et al. Automatic Construction of the Initial Orbitals for Efficient Generalized Valence Bond Calculations of Large Systems. *J. Chem. Theory. Comput.* **2019,** *15*, 141-153.
(15) Carter, E. A.; Goddard, W. A. Correlation‐consistent configuration interaction: Accurate bond dissociation energies from simple wave functions. *J. Chem. Phys.* **1988,** *88*, 3132-3140.
(16) Murphy, R. B.; Pollard, W. T.; Friesner, R. A. Pseudospectral localized generalized Mo/ller–Plesset methods with a generalized valence bond reference wave function: Theory and calculation of conformational energies. *J. Chem. Phys.* **1997,** *106*, 5073-5084.
(17) Lawler, K. V.; Beran, G. J.; Head-Gordon, M. Symmetry breaking in benzene and larger aromatic molecules within generalized valence bond coupled cluster methods. *J. Chem. Phys.* **2008,** *128*, 024107.
(18) Kutzelnigg, W. Separation of strong (bond-breaking) from weak (dynamical) correlation. *Chem. Phys.* **2012,** *401*, 119-124.
(19) Zoboki, T.; Szabados, A.; Surjan, P. R. Linearized Coupled Cluster Corrections to Antisymmetrized Product of Strongly Orthogonal Geminals: Role of Dispersive Interactions. *J. Chem. Theory. Comput.* **2013,** *9*, 2602-8.
(20) Pernal, K. Intergeminal Correction to the Antisymmetrized Product of Strongly Orthogonal Geminals Derived from the Extended Random Phase Approximation. *J. Chem. Theory. Comput.* **2014,** *10*, 4332-41.
(21) Li, S. Block-correlated coupled cluster theory: the general formulation and its application to the antiferromagnetic Heisenberg model. *J. Chem. Phys.* **2004,** *120*, 5017-26.
(22) Fang, T.; Li, S. Block correlated coupled cluster theory with a complete active-space self-consistent-field reference function: the formulation and test applications for single bond breaking. *J. Chem. Phys.* **2007,** *127*, 204108.
(23) Frisch, M. J., et al. *Gaussian 16 Rev. A.03*, Wallingford, CT, 2016.
(24) Schmidt, M. W., et al. General atomic and molecular electronic structure system. *J. Comput. Chem.* **1993,** *14*, 1347-1363.
(25) Sun, Q., et al. PySCF: the Python-based simulations of chemistry framework. *WIREs Comput Mol Sci.* **2018,** *8*, e1340.